\documentclass[11pt]{article}
\usepackage{fleqn,cosparb}

\usepackage{url}

\usepackage{graphicx}
\usepackage[figuresright]{rotating}

\title{AN UNEXPLAINED 10${}^\circ$-40${}^\circ$ SHIFT IN THE LOCATION OF SOME DIVERSE NEUTRAL ATOM DATA AT 1~AU}

\author{Michael R. Collier\address{NASA/Goddard Space Flight Center, 
    Greenbelt, MD 20771},
        Thomas E. Moore$^{1}$,
        David Simpson${}^{1}$, Aaron Roberts${}^{1}$, Adam Szabo${}^{1}$,
        Stephen Fuselier\address{Lockheed Martin Advanced Technology Center, 
        Palo Alto, CA 94304},
        Peter Wurz\address{University of Bern, CH-3012, Switzerland},
        Martin A. Lee\address{University of New Hampshire, Durham 03824}, 
        and Bruce T. Tsurutani\address{Jet Propulsion Laboratory,
        California Institute of Technology, Pasadena 91109}}

\begin{document}
\voffset 0.25in

\maketitle

\begin{abstract}
Four different data sets pertaining to the neutral atom environment at 1~AU are presented and discussed. These data sets include neutral solar wind and interstellar neutral atom data from IMAGE/LENA, energetic hydrogen atom data from SOHO/HSTOF and plasma wave data from the magnetometer on ISEE-3. Surprisingly, these data
sets are centered between 262${}^\circ$ and 292${}^\circ$ ecliptic longitude, $\sim$10${}^\circ$-40${}^\circ$ from the upstream interstellar neutral (ISN)
flow direction at 254${}^\circ$ resulting from the motion of the Sun relative to the local interstellar cloud (LIC). Some possible explanations for this offset, none of which is completely satisfactory, are discussed.
\end{abstract}

\section*{INTRODUCTION}
\vskip10pt

Due to the motion of the heliosphere at about 25~km/s through the LIC, the Earth passes upstream of the Sun in the main neutral gas flow in early June of every year, about June 5 (day 156) when it is near 254${}^\circ$ ecliptic longitude
(Frisch, 2000). Several independent observations for both H and He including 
pickup ions (Gloeckler and Geiss, 2001), direct neutral gas observations (Witte et al., 1993), and UV backscattering (Lallement, 1996) have established this
direction along with the resulting spatial distribution and kinematics of the
particles. In addition, the derived flow is consistent with UV absorption 
measurements in the light of nearby stars (Bertin et al., 1993).

The presence of this well-established stream leads to the expectation that
neutral atom data at 1~AU would be symmetric with respect to the 74${}^\circ$/254${}^\circ$ ecliptic longitude axis. However, a number of
neutral atom data sets at 1~AU, four of which are discussed here, curiously
are not centered with this axis, but with larger ecliptic longitudes by about
10${}^\circ$-40${}^\circ$, depending on the data set in question.

\begin{table}[t]
\begin{minipage}{105mm}
\includegraphics[width=105mm]{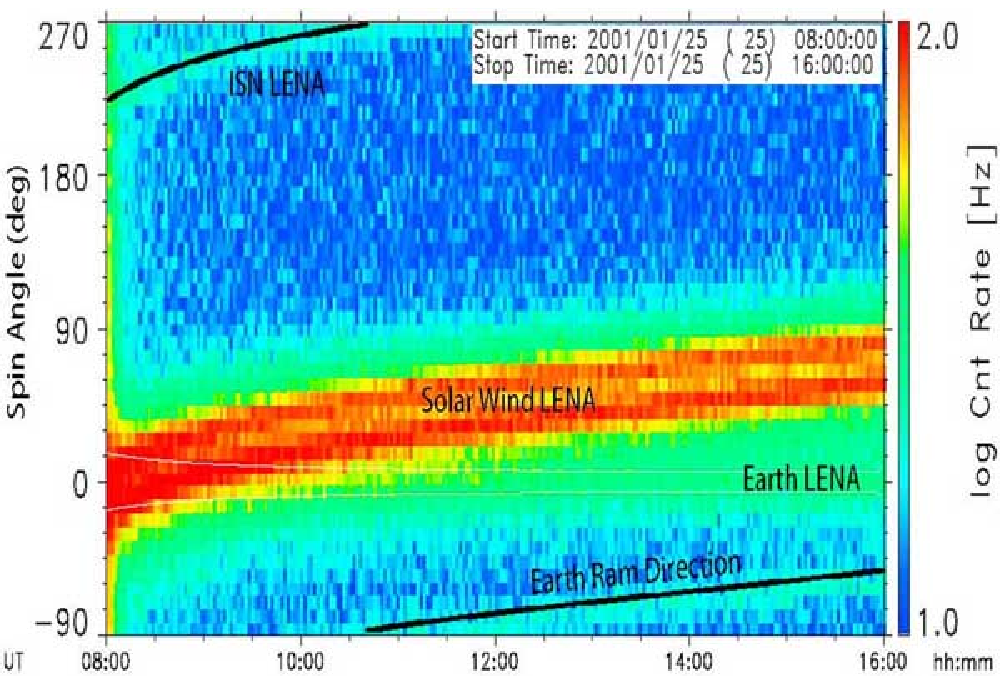}
{\sf Fig. 1. LENA spectrogram showing ISN signal.}
\end{minipage}
\hfil\hspace{\fill}
%
\begin{minipage}{80mm}
  \includegraphics[width=80mm]{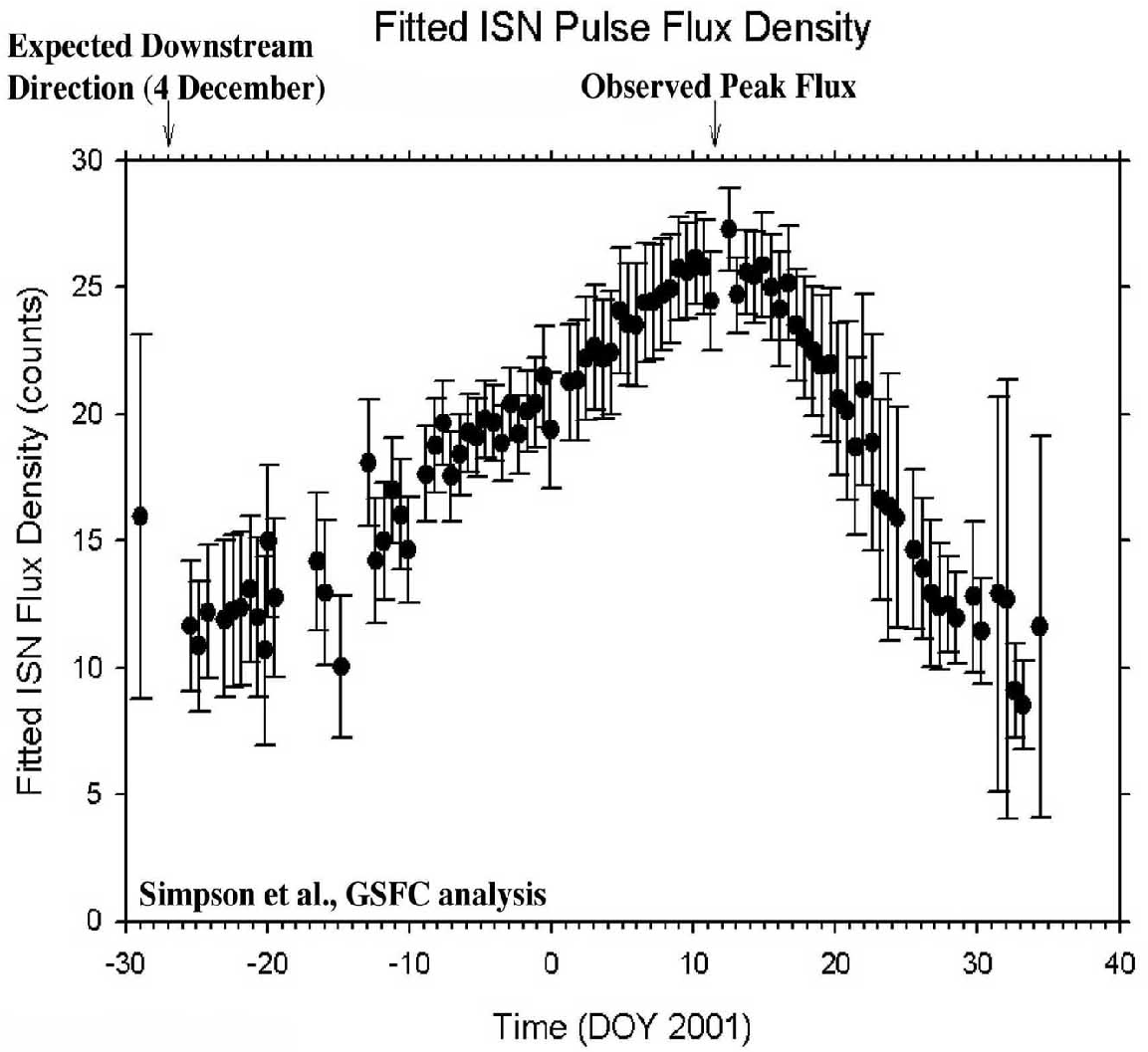}
{\sf Fig. 2. ISN count rate versus day of year.}
\end{minipage}
\end{table}%

\section*{INTERSTELLAR NEUTRAL (ISN) OBSERVATIONS}
\vskip10pt

Fuselier predicted prior to the IMAGE launch in March of 2000, based in part on earlier unpublished work by Gruntman, that the Low Energy Neutral Atom (LENA)
imager, which responds to neutral atoms down to as low as about 10~eV (Moore et al., 2000), would be able to directly observe interstellar neutral helium early
in each calendar year. As shown in Figure~1, LENA did observe a signal in the Winter of 2000/2001 which, because it occurred at the predicted time of year and from the predicted direction, close to the Earth ram direction, was interpreted
as due to ISN.

Because the upstream direction, 254${}^\circ$, lies outside of LENA's field-of-view, for ISN to be observed directly by LENA they must be appreciably bent by the Sun's gravity downwind of the Sun, making it unlikely that this signal is ISN hydrogen which is strongly influenced by solar radiation pressure. Consequently, the signal is probably helium, although LENA does respond to all species of neutrals over a wide energy range. 

The peak interstellar neutral flux is expected when the Earth is directly downstream, on December 5. However, because the LENA efficiencies are a strong
function of energy and the expected velocity of heavy ISN with respect to IMAGE
exhibits a broad peak starting in mid-December, the maximum in the observed neutral count rate should occur somewhere around December 15. So, following the
appearance of this signal, two groups on the LENA team independently used different techniques to extract the signal and track its rate versus time. The
two groups reached the same conclusion, namely that the peak count rate of
the neutrals occurred about forty days later than December 5, in early January,
as shown in Figure~2.
Thus, if of ISN origin, these observations do not seem to come from the same population of neutrals observed by the Ulysses Neutral GAS experiment (Witte et al., 1992), through UV backscattering (Chassefi\`ere et al., 1986; Vallerga, 1996) and through pickup ions (M\"obius et al., 1985; Gloeckler and Geiss, 2001).

\section*{NEUTRAL SOLAR WIND (NSW) OBSERVATIONS}
\vskip10pt

Figure~3 shows the annual variation of the neutral solar wind flux, which forms
when solar wind ions exchange charge with neutral atoms between the Sun and the
Earth (Collier et al., 2001), observed by LENA over the year 2001 (dashed line).
There is a clear enhancement of about one and one half orders of magnitude
in the data occurring between about day 120 and day 250, although the Sun, and
hence the solar wind, is outside of LENA's field-of-view from about day 144 to
day 230. When the center of the enhancement is inferred from the rise and fall,
its location is estimated to be on day 184.

The highest measured neutral solar wind flux is $\sim0.2\%$ of the nominal
solar wind flux, although the upward trend may suggest a higher peak rate. 
The flux is based on the assumption that LENA responds to hydrogen with an
energy of 1~keV. However, the average solar wind energy is higher than 1~keV
and LENA's efficiency may be higher at the higher energies. Furthermore, LENA
may be responding in-part to suprathermal particles or heavy atoms, which also
will have higher efficiencies. Considering this and uncertainties in 
calibration, the observed neutral solar wind flux could be lower than that
shown in Figure~3 by about a factor of four or five. Because IMAGE, except
under infrequent extreme conditions, is in the magnetosphere, there are no
solar wind ions to suppress.

The neutral solar wind fluxes outside of the period of enhancement have been interpreted as neutrals generated by the solar wind interaction with interplanetary dust and the Earth's hydrogen exosphere (Collier et al., 2002).
The third main source of neutrals for solar wind charge exchange, interstellar
hydrogen, which has a higher density and will charge exchange more readily with
protons than neutral helium will (Gruntman, 1994), is expected to have an
annual periodicity due to the Earth's motion around the Sun. Furthermore,
unlike helium, hydrogen is relatively unaffected by solar gravity, being
partially if not entirely balanced by radiation pressure, so that the highest
hydrogen densities are found in the upstream, rather than downstream, region.
Recent evidence with SOHO/SWAN indicates a rather high photon pressure with $\mu$, the ratio of the radiation to gravitational force, approaching one even during solar minimum (Qu\'emerais et al., 1999).

Figure~6 of Bzowski et al. (1996) shows a model prediction for the annual
variation of the neutral solar wind flux at 1~AU over a solar cycle.
Qualitatively, the model predicts a variation between one and three orders
of magnitude in the upstream direction, consistent with the LENA observations
and prompting an interpretation linking the LENA enhancement with interstellar hydrogen. However, the observed fluxes are over an order of magnitude greater
than the predicted fluxes, which are close to 10${}^4$~atoms/cm${}^2$/s, and
occur about thirty days (or approximately 30${}^\circ$) later than the
nominal interstellar neutral flow direction.

\begin{table}[t]
%
\begin{minipage}{80mm}
  \includegraphics[width=80mm]{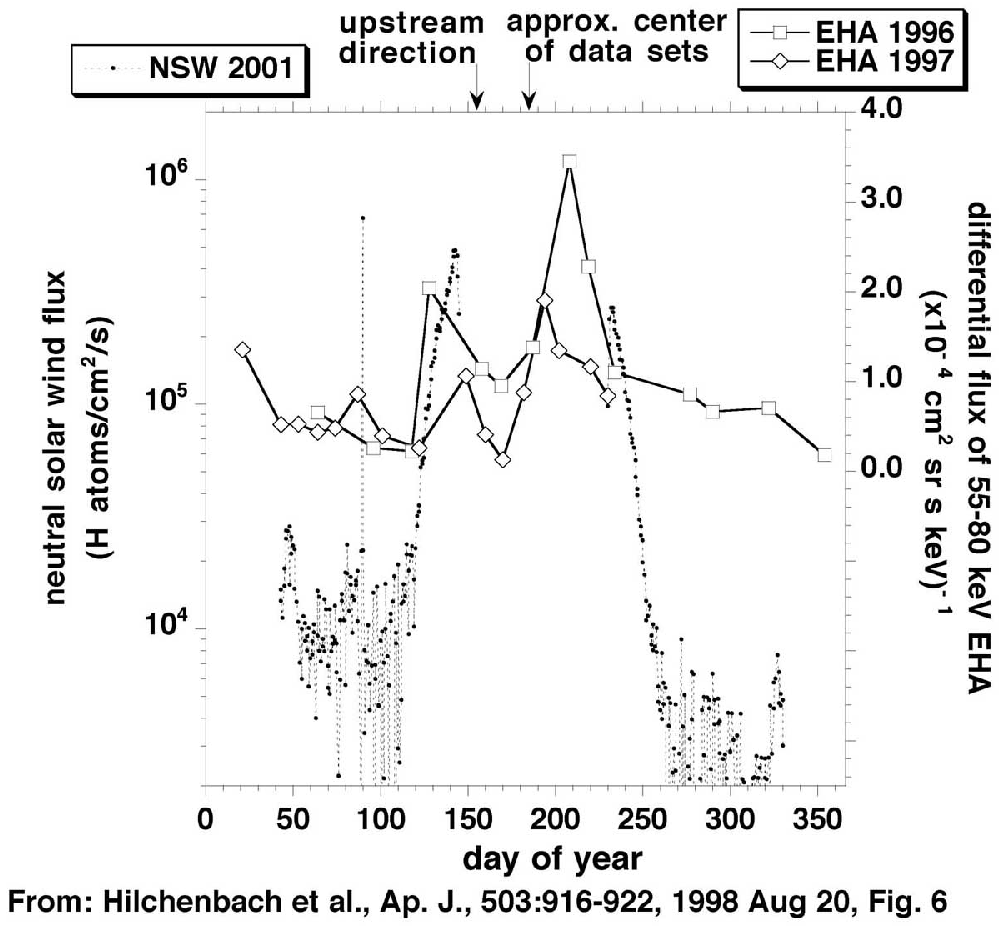}
{\sf Fig. 3. HSTOF/EHA and LENA/NSW data.}
\end{minipage}
%
\hskip 2cm
\begin{minipage}{65mm}
\includegraphics[width=65mm]{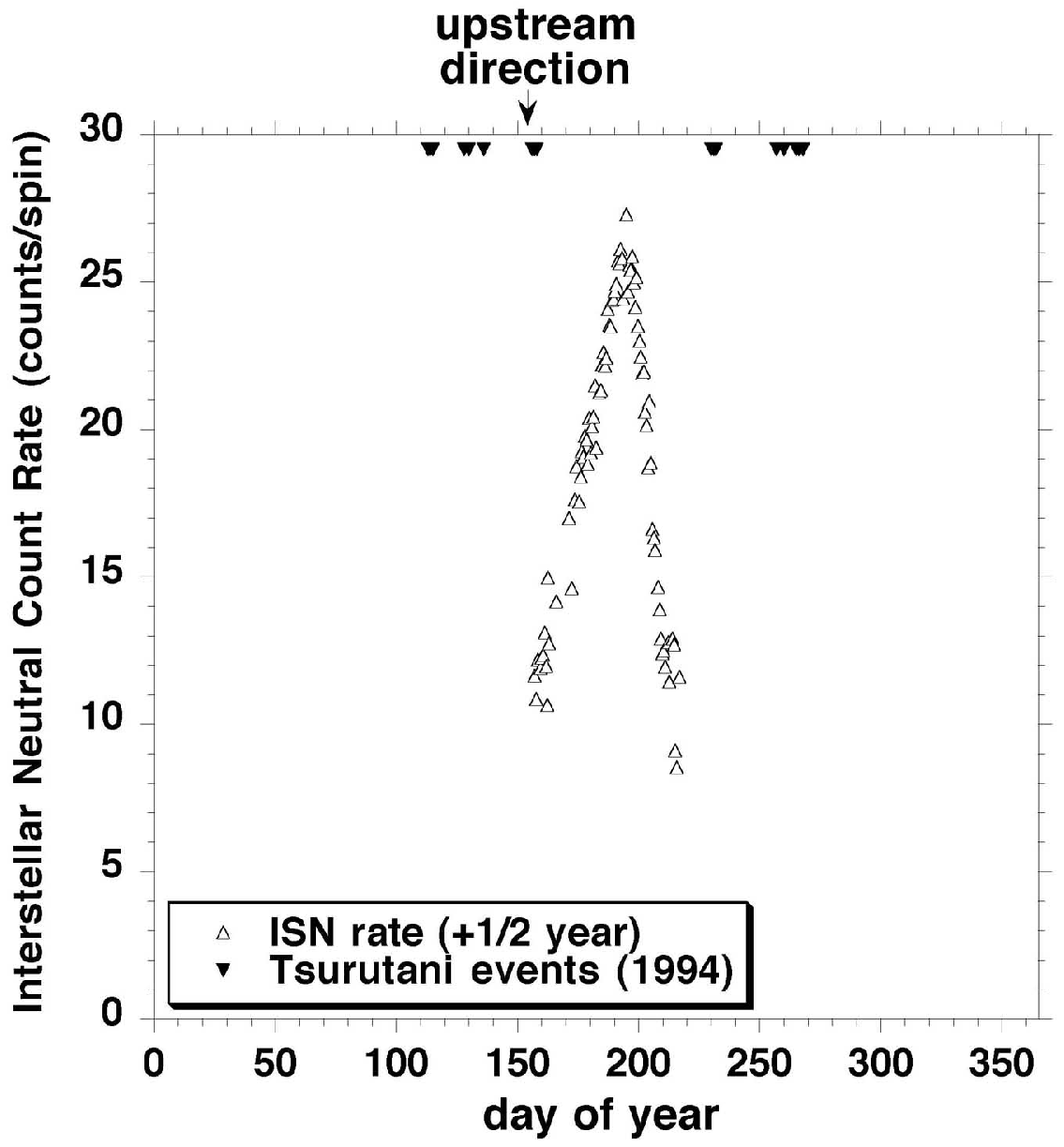}
{\sf Fig. 4. Tsurutani events and ISN data (shifted 6 months).}
\hfil\hspace{\fill}
\end{minipage}
\end{table}%


\section*{ENERGETIC HYDROGEN ATOM OBSERVATIONS}
\vskip10pt

Figure~3 also shows energetic hydrogen atom data (solid line) from the High Energy Suprathermal Time-of-Flight (HSTOF) sensor on SOHO published by Hilchenbach et al. (1998)
(data from their Figure~6a). They examined quiet day fluxes of hydrogen atoms with energies
between 55 and 80~keV and interpret these fluxes as coming from the heliosheath.
K\'ota et al. (2001) have argued that the HSTOF ENA observations are also consistent with an energetic ion population source accelerated at CIRs in the inner heliosphere.

Like the LENA neutral solar wind observations, HSTOF is looking back towards
the Sun and, like LENA, HSTOF sees an enhancement in energetic neutrals
between about day 120 and 250. However, unlike LENA, HSTOF is not looking
directly back at the Sun, but 37${}^\circ$ off the Sun-Earth line.
When the data are plotted as a function of the actual ecliptic
longitude HSTOF observes (see Hilchenbach et al., Figure~6b), there is a
substantial shift, about 15${}^\circ$, between the peak flux and the
nominal upstream/downstream axis, although the statistical uncertainties
are relatively large.
This shift is apparent in the HSTOF long term trending data shown in Figure~18 of Czechowski et al. (2001) as well. However, the fluxes HSTOF observes are higher by an order of magnitude than can be accounted for by the models considered by Czechowski et al. Certainly an additional source of neutral gas, such as might be supplied by a secondary stream, would bring model and observations into closer agreement.


\section*{WAVE OBSERVATIONS}
\vskip10pt

Tsurutani et al. (1994) reported low frequency waves with periods near the proton gyroperiod at 1~AU observed by the
magnetometer on ISEE-3. The events are unusual because
the interplanetary magnetic field power spectrum at 1~AU
is typically quite featureless, exhibiting a relatively smooth Kolmogorov
$\nu^{-5/3}$ dependence. However, during these events
(see their Figure~3), Tsurutani et al. saw broad increases in the wave power near the proton cyclotron frequency, atypical in the normal solar wind.

Tsurutani et al. considered pickup of cold hydrogen neutrals as the most likely source of the waves and list interstellar neutrals as a possible candidate.
The dates of their events are distributed over a three year period from 1978-1981 (see their Table 1). However, the day of year of these events falls into two clusters, as shown at the top of Figure~4, which appear to be centered not with the upstream direction, but about thirty degrees later.

In the event these wave observations are associated with elevated neutral fluxes
centered at an ecliptic longitude somewhere between 262${}^\circ$ and
292${}^\circ$, then a natural question is why would this wave activity only occur in the regions of the 
neutral atom gradients. One possible explanation is that
it results from Earth crossings of the parabolic exclusion boundary (Holzer, 1977). For values of $\mu>1$, hydrogen is unable to penetrate to the Sun and the
fobidden region forms a parabolic boundary, which, in analogy to the magnetosheath, has an associated hydrogen sheath of substantially increased density. For a static boundary and reasonable values of $\mu$, the Earth will traverse this sheath twice annually in the upstream direction (see Holzer, 1977, Figure~4b). However, the boundary is likely irregular and in near constant motion, causing multiple traversals and bursty activity during the appropriate times of year,
as observed in Tsurutani et al.'s events. In fact, examining Tsurutani et al.'s wave events, they do resemble, in the sense of having multiple closely-spaced
events, the traversals of boundaries such as the magnetopause and bow shock.

Of course, if the secondary stream population is very hot, only those particles in the distribution with energies high enough to effectively penetrate to 1~AU while low enough to form a parabolic exclusion boundary near 1~AU would be producing these waves. Note also that Holzer uses a value $\mu$=1.2, whereas
$\mu$ may be substantially larger leading to higher energy particles forming the same parabolic exclusion boundary.

\section*{DISCUSSION AND CONCLUSIONS}
\vskip10pt

Figure~5 shows all four of the data sets discussed in this paper on a single plot. The data have a symmetry point substantially later than 
expected based on the nominal upstream direction
but appear to be consistent with a
direction very close to 
the Galactic center at 267${}^\circ$. One possibility
is that this may be due to a secondary stream of neutrals 
which enters the heliosphere at an ecliptic longitude 
somewhere between 262${}^\circ$ and 292${}^\circ$.
The wave data and the downstream directly observed neutral data suggest a lower
energy component while the neutral solar wind and perhaps the HSTOF data favor
a component at higher energies which can penetrate well inside of 1~AU. This 
implies that should this secondary stream exist, it likely contains a wide range of neutral speeds, that is, it is very hot.

A natural question is what would cause such a stream and the answer
is unclear at best, although
there are a couple other relevant issues that should be mentioned. 
First, it is interesting to note that the apparent direction of the interstellar
dust flow is shifted about 10${}^\circ$ later in ecliptic longitude than
the direction of the interstellar neutral flow (Gr\"un, 2000), although they
are consistent to within a 1$\sigma$ uncertainty. The dust distribution,
however, is sufficiently broad so that it is also consistent with a wide
range of flow directions.
Because dust can serve as a source of neutrals for charge exchange
(Banks, 1971), there
may be some relationship between the interstellar dust flow and this 
possible secondary stream.

Second, 
if the heliosphere is tilted due to the inclination of an interstellar magnetic field, as suggested by some simulations (Ratkiewicz et al., 1998)
and illustrated in Figure~6, 
then perhaps the shift in the data sets presented here is the result of this
asymmetry.

Third, Lallement (private communication, 2002) has pointed out that evidence suggests that the
heliosphere is extremely close to the boundary of the local interstellar
cloud in the approximate direction of the Galactic center, albeit with
a huge uncertainty
(Lallement, 1996; Lallement and Bertin, 1992). If the next cloud in that
direction, the G cloud (Linsky and Wood, 1996), has not already caught up to the
LIC (Lallement et al., 1990), between the LIC and the G cloud resides a hot ionized gas of temperature $\sim$10${}^6$~K, which corresponds to 150~km/s for protons. If the interface with this hot gas is close (we have only upper limits (Redfield and Linsky, 2000)), then, because of charge exchange between hydrogen and protons, hot neutral H with characteristic speeds of about 150~km/s, perhaps
higher, will penetrate the heliosphere from the approximate direction of the Galactic center because of our proximity to the interface.
In this scenario, because the distribution is so hot, {\it some\/} flux of energetic neutrals ($>1$~keV) may be observed flowing 
{\it towards} the Sun when the Earth is between the Sun and the Galactic center.

Fourth, the role of the Earth in influencing 1~AU observations of extraterrestrial neutrals may not be fully appreciated. As just one example, because lower energy particles in the interstellar neutral distribution are preferentially filtered out, the remnant population at 1~AU will have a higher effective speed. The Earth moves at ${\rm v}_{\rm E}$=30~km/s. Thus, if the interstellar neutrals flow at ${\rm v}_{\rm ISN}$ and
the Earth is at an angle $\sin{}^{-1}\{{\rm v}_{\rm E}/{\rm v}_{\rm ISN}\}$ with respect to the upstream direction, then the interstellar neutrals
will flow exactly along the Sun-Earth line in the frame of the Earth. 
Consequently, the axis of
the focusing cone created by the Earth's gravity will be aligned with the
solar wind and will produce strong neutral density enhancements along an extended solar wind path length.

Fifth, a closer look at the spatial distribution of pickup ions appears 
to be warranted (Gloeckler and Geiss, 2001), with a shorter averaging window as used, for example, by M\"obius et
al. (2002). This treatment appears to reveal evidence of primary and secondary streams during the period of the IMAGE data sets (2000-2001). However, it must be borne in mind that the spectral form of the pickup ions resultant from a fast
and/or warm secondary stream may not have the same form as that which results 
from cold interstellar neutral photoionization. The spectral form that is a 
cut-off at two times the solar wind velocity results from pickup of 
neutrals travelling slowly with respect to the solar wind.

In summary, 
multiple data sets exhibit a spatial structure that is aligned with a
direction between 
10${}^\circ$ and 40${}^\circ$ from the nominal upwind direction of the 
interstellar flow.
This structure may possibly be explained in terms of a secondary stream.
It remains to be fully understood, however, how these data fit in with previous measurements of neutral atoms, pickup ions and UV spectra within the
heliosphere and whether or not this interpretation is consistent with these observations.

\begin{table}[t]
%
\begin{minipage}{87mm}
\includegraphics[width=87mm]{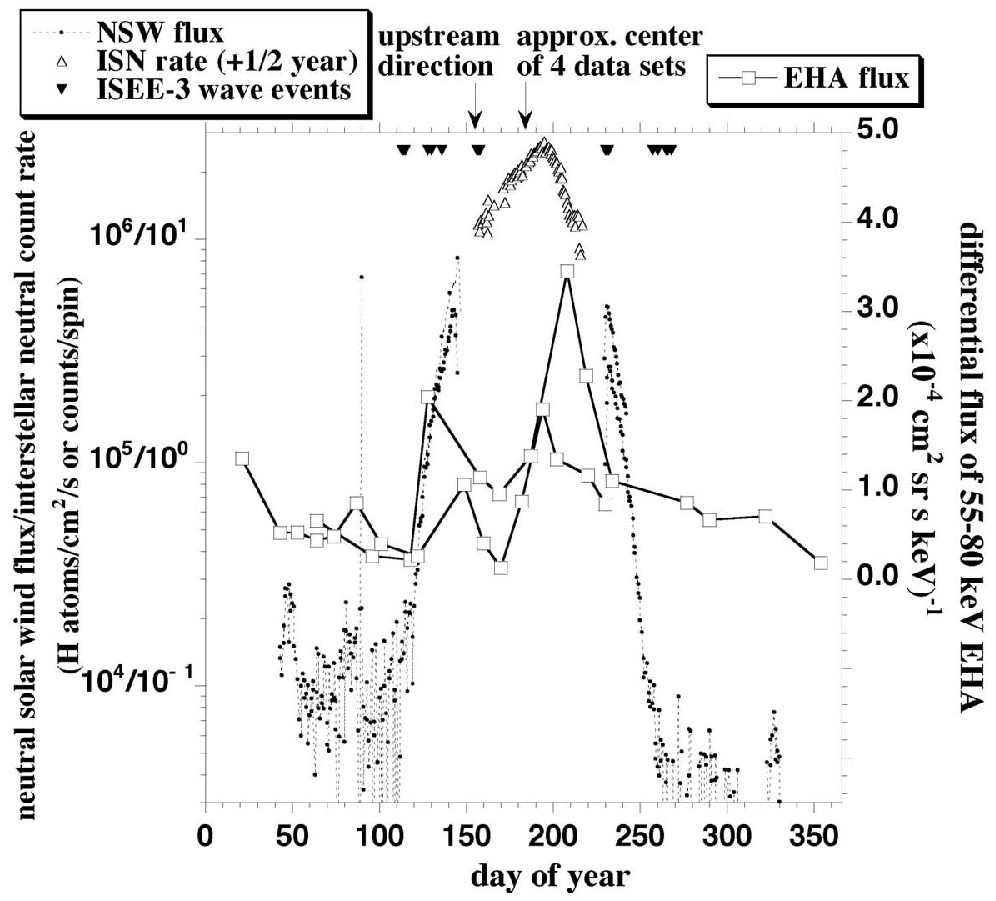}
{\sf Fig. 5. Four data sets discussed here (ISN data shifted 6 months).}
\hfil\hspace{\fill}
\end{minipage}
%
\hskip 0.15cm
\begin{minipage}{97mm}
\includegraphics[width=97mm]{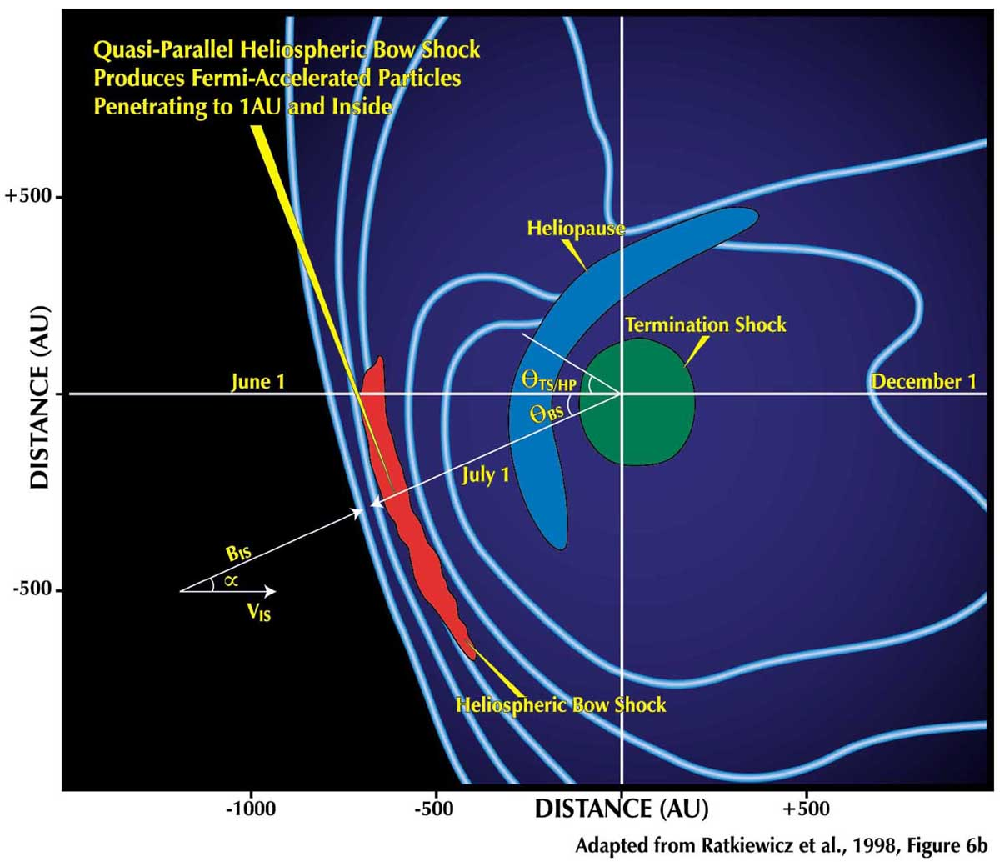}
{\sf Fig. 6. Angled IS B-field producing heliospheric tilt.}
\hfil\hspace{\fill}
\end{minipage}
\end{table}%

\section*{ACKNOWLEDGMENTS}
\vskip10pt

Special thanks to R. Lallement and J-L Bertaux for valuable discussions.
Portions of this research were performed at the Jet Propulsion Laboratory, Calif. Inst. of Tech. under contract with NASA.

\vskip10pt
\noindent
E-mail address of Michael R. Collier\hskip20pt \underbar{mcollier@pop600.gsfc.nasa.gov}
\vskip0pt \noindent
Manuscript received 28 November 2002; revised 31 March 2003, accepted 31 March 2003

\end{document}